\documentclass[iop]{emulateapj}

\newcommand{\integral}{{\it INTEGRAL}}
\newcommand{\suzaku}{{\it Suzaku}}
\newcommand{\swift}{{\it Swift}}
\newcommand{\xmm}{{\it XMM-Newton}}
\newcommand{\asca}{{\it ASCA}}

\shorttitle{\textit{SUZAKU} VIEW OF THE \textit{SWIFT}/BAT ACTIVE GALACTIC NUCLEI. III.}
\shortauthors{Eguchi et al.}

\begin{document}

\submitted{Accepted on December 30, 2010}

\title{
\textit{SUZAKU} VIEW OF THE \textit{SWIFT}/BAT ACTIVE GALACTIC
NUCLEI. III.  APPLICATION OF NUMERICAL TORUS MODELS TO TWO NEARLY
COMPTON THICK AGNS (NGC~612 AND NGC~3081)
}

\author{
 Satoshi Eguchi\altaffilmark{1},
 Yoshihiro Ueda\altaffilmark{1},
 Hisamitsu Awaki\altaffilmark{2},
 James Aird\altaffilmark{3},
 Yuichi Terashima\altaffilmark{2},
 and Richard Mushotzky\altaffilmark{4}
}

\altaffiltext{1}{Department of Astronomy, Kyoto University, Kyoto 606-8502, Japan}
\altaffiltext{2}{Department of Physics, Faculty of Science, Ehime University, Matsuyama 790-8577, Japan}
\altaffiltext{3}{Center for Astrophysics and Space Sciences (CASS), Department of Physics, University of California, San Diego, CA 92093, USA}
\altaffiltext{4}{Department of Astronomy, University of Maryland, College Park, MD, USA}

\begin{abstract}

The broad band spectra of two {\swift}/BAT AGNs obtained from
{\suzaku} follow-up observations are studied: NGC~612 and NGC~3081.
Fitting with standard models, we find that both sources show similar
spectra characterized by a heavy absorption with $N_{\rm{H}} \simeq
10^{24} \ \rm{cm}^{-2}$, the fraction of scattered light is
$f_{\rm{scat}} = 0.5-0.8\%$, and the solid angle of the reflection
component is $\Omega/2\pi = 0.4-1.1$. To investigate the geometry of
the torus, we apply numerical spectral models utilizing Monte Carlo
simulations by \citet{Ikeda2009} to the {\suzaku} spectra. We find our
data are well explained by this torus model, which has 
four geometrical parameters. The fit results suggest
that NGC~612 has the torus half opening-angle of $\simeq
60^{\circ}-70^{\circ}$ and is observed from a nearly edge-on angle
with a small amount of scattering gas, while NGC~3081 has a very small
opening angle $\simeq 15^\circ$ and is observed on a face-on geometry,
more like the deeply buried ``new type'' AGNs found by
\citet{Ueda2007}. We demonstrate the potential power of direct
application of such numerical simulations to the high quality broad
band spectra to unveil the inner structure of AGNs.

\end{abstract}


\section{Introduction}

The strong correlation between the mass of a supermassive black hole (SMBH)
and that of the galactic bulge \citep[e.g.,][]{Magorrian1998,
Marconi-Hunt2003} suggests a fundamental link between the growth of SMBH
and galaxy evolution. Theoretical models predict that most SMBHs in
galaxies experience a heavily obscured phase in their growth stage
\citep[e.g.,][]{Hopkins2005}. Indeed, studies based on population
synthesis models of the cosmic X-ray background (CXB) suggest that
heavily obscured AGNs, whose line-of-sight hydrogen column density
($N_{\rm{H}}$) is greater than $10^{23.5} \ \rm{cm}^{-2}$, are a significant
fraction of the AGN population \citep{Ueda2003, Gilli2007}. Due to the
difficulty of detecting them in most  energy bands, however, our
understanding of heavily obscured AGNs (including ``Compton-thick''
ones with $N_{\rm H} > 10^{24} \ \rm{cm}^{-2}$) is very scarce even in
the local universe.

Sensitive hard X-ray observations above 10 keV, where the penetrating
power overwhelms photo-electric absorption, provide fruitful
information about this population, except for heavily Compton thick
($N_{\rm{H}} \gtrsim 10^{24.5} \ \rm{cm}^{-2}$) objects.  Recent all
sky hard X-ray surveys performed with {\swift}/BAT (15--200 keV;
\citealt{Tueller2008}) and {\integral} (10--100 keV;
\citealt{Bassani2006, Krivonos2007}) are ideal for this purpose with
much less selection biases than surveys at lower energies.

\begin{deluxetable*}{cccccc}
\tabletypesize{\scriptsize}
\tablecaption{List of Targets\label{tab-targets}}
\tablewidth{0pt}
\tablehead{ \colhead{SWIFT} & \colhead{Optical/IR Identification}
 & \colhead{R.A. (J2000)} & \colhead{Dec. (J2000)}
 & \colhead{Redshift} & \colhead{Classification} }
\startdata
 J0134.1--3625 & NGC 612 & 01 33 57.74 & -36 29 35.7 & 0.0298 & Seyfert 2 \\
 J0959.5--2258 & NGC 3081 & 09 59 29.54 & -22 49 34.6 & 0.0080 & Seyfert 2 \\
\enddata
\tablecomments{The position, redshift, and classification for each source is taken from the NASA/IPAC Extragalactic Database.}
\end{deluxetable*} 

Our team have been working on a systematic follow-up observation
program with {\suzaku} of Swift/BAT detected AGNs whose broad band
X-ray spectra were poorly (or never) studied previously, targeting
obscured objects in most cases. \citet{Ueda2007} discovered deeply
buried AGNs that exhibit very small fractions of scattered soft X-rays
($<0.5 \%$) with respect to the transmitted component, with strong
reflection signals most probably coming from the inner wall of the
Compton-thick tori. Further studies of six Swift AGNs by
\citet{Eguchi2009} (Paper I hereafter) show that they could be
classified into two types, ``new type'' AGNs with a small scattering
fraction and strong reflection strength, and ``classical type'' ones
with a larger scattering fraction and weaker reflection.  These types
are consistent with SMBHs surrounded by geometrically thick and thin
tori, respectively. Using an {\integral} selected
sample, \citet{Comastri2009} also suggest that there are distinct
AGN populations of new and classical types, although the result
depends on whether the absorption for the reflection component is
considered or not in the spectral model \citep[see][]{Comastri2010}.
Due to the limited number of objects in the sample studied we are far
from reaching a consensus on the torus structure and its dependence on
various parameters like the AGN luminosity, Eddington ratio, and
properties of the host-galaxy, for the whole AGN populations.

High quality broad band X-ray spectra give unique insight into the
structure and geometry of the central region of AGNs. Most previous
studies, however, relied on phenomenological spectral models where the
detailed geometry of the torus is not taken into account; usually, an
analytical formula for the Compton reflection from matter with
infinite optical depths is simply assumed for the reprocessed
emission, and absorption column density of the transmitted component
is treated independently. Monte Carlo simulation is a powerful tool to
reproduce realistic spectra from AGNs with a complex structure of the
torus, which may not always have a sufficiently large optical depth
for Compton scattering. Recently, \citet{Ikeda2009} have developed
such a Monte Carlo code that can be applicable to the broad band X-ray
spectra with several free parameters describing the torus
geometry. Similarly, \citet{Murphy2009} also studied
the numerical spectra from a toroidal torus, known as the MYTORUS
model\footnote{\texttt{http://www.mytorus.com/}}, although we do not
adopt this model here because the opening angle of the torus is fixed
at 60$^\circ$. Applying such models directly to the observed spectra,
we can obtain more accurate constraints on the inner structure of AGNs
than from the standard previous analysis.

In this paper, we present the results of detailed X-ray spectral
analysis of {\suzaku} data of two {\swift}/BAT AGNs, Swift
J0134.1--3625 (NGC~612; $z = 0.0298$) and
Swift J0959.5--2258 (NGC~3081; $z = 0.0080$),
whose simultaneous broad band spectra were not available before.
NGC~612 is a powerful radio galaxy, which was originally classified as
Fanaroff-Riley (FR; \citealt{Fanaroff1974}) II type by
\citet{Morganti1993}. This object hosts prominent double radio
sources; the eastern lobe has a bright hot spot near its outer edge,
while the western one has a jet-like structure. Since the former and
latter morphology correspond to those of the FR I and II types,
respectively, \citet{Gopal2000} classifies it as a hybrid morphology
radio source.
NGC~612 shows an optical spectrum of Seyfert 2 galaxies but the
intensity of the [\ion{O}{3}] emission is very weak
\citep{Parisi2009}.
\citet{Winter2008} present the X-ray spectrum observed with {\xmm},
obtaining a large hydrogen column density of $N_{\rm H} \simeq 10^{23.9} \
\rm{cm}^{-2}$ with an apparently very flat power-law index of
$\simeq$0.3, suggestive of a reflection-dominant spectrum below 10
keV.
NGC~3081 is a Seyfert 2 hosted by a barred galaxy. This object has
three rings associated with a tidal interaction \citep{Freeman2000},
and exhibits time variable polarization in the optical band
\citep{Joshi1989}. The strong [\ion{O}{3}] emission is
observed from this object \citep{Storchi-Bergmann1995}.
\citet{Moran2001} report a large absorption column density ($N_{\rm H}
\simeq 10^{23.7} \ \rm{cm}^{-2}$) from the X-ray spectrum observed
with {\asca}.

Section~\ref{sec-observation} describes the observations and data reduction of the two sources.
We first analyze the spectra with standard spectral models in
Section~\ref{sec-analytical-model}, and then present the results of
application of the torus model utilizing Monte Carlo calculation by
\citet{Ikeda2009} in Section~\ref{sec-torus-model}. The implications of
our results are discussed in Section~\ref{sec-discussion}.
We adopt the cosmological parameters
($H_{0}$, $\Omega_{\rm{m}}$, $\Omega_{\rm{\lambda}}$) =
($70 \ \rm{km} \ \rm{s}^{-1} \ \rm{Mpc}^{-1}$, $0.3$, $0.7$)
throughout the paper.

\section{Observation and Data Reduction}\label{sec-observation}

\subsection{Observation}

We observed NGC~612 and NGC~3081 with {\suzaku} in 2008 May and June,
respectively.
The basic information for our targets is summarized in
Table \ref{tab-targets}. {\suzaku} \citep{Mitsuda2007} carries four
X-ray CCD cameras called the X-ray Imaging Spectrometer (XIS-0, XIS-1,
XIS-2, and XIS-3) as focal plane imager of four X-ray telescopes, and
a non-imaging instrument called the Hard X-ray Detector (HXD)
consisting of Si PIN photo-diodes and GSO scintillation
counters. XIS-0, XIS-2, and XIS-3 are front-side illuminated CCDs
(FI-XISs), while XIS-1 is the back-side illuminated one (BI-XIS).
To maximize the effective area of the HXD, 
the targets were observed at the HXD nominal position, which is
about 5 arcmin off-axis from the averaged optical axis of the
XISs.\footnote{\texttt{http://heasarc.gsfc.nasa.gov/docs/suzaku/analysis/abc/
}}

\begin{deluxetable*}{cccccc}
\tabletypesize{\scriptsize}
\tablecaption{Observation Log\label{tab-observations}}
\tablewidth{0pt}
\tablehead{ \colhead{Target} & \colhead{Start Time (UT)} & \colhead{End Time}
 & \colhead{Exposure\tablenotemark{a} (XIS)} & \colhead{Exposure (HXD/PIN)} & \colhead{SCI\tablenotemark{b}} }
\startdata
NGC~612 & 2008 May 20 16:19 & May 21 20:08 & 48.5 ks & 41.3 ks & On \\
NGC~3081 & 2008 Jun 18 21:49 & Jun 19 19:33 & 43.7 ks & 42.4 ks & On \\
\enddata
\tablenotetext{a}{Based on the good time interval for XIS-0.}
\tablenotetext{b}{With/without the spaced-row charge injection for the XIS \citep{Nakajima2008}.}
\end{deluxetable*}

\begin{figure*}
\epsscale{1.0}
\plotone{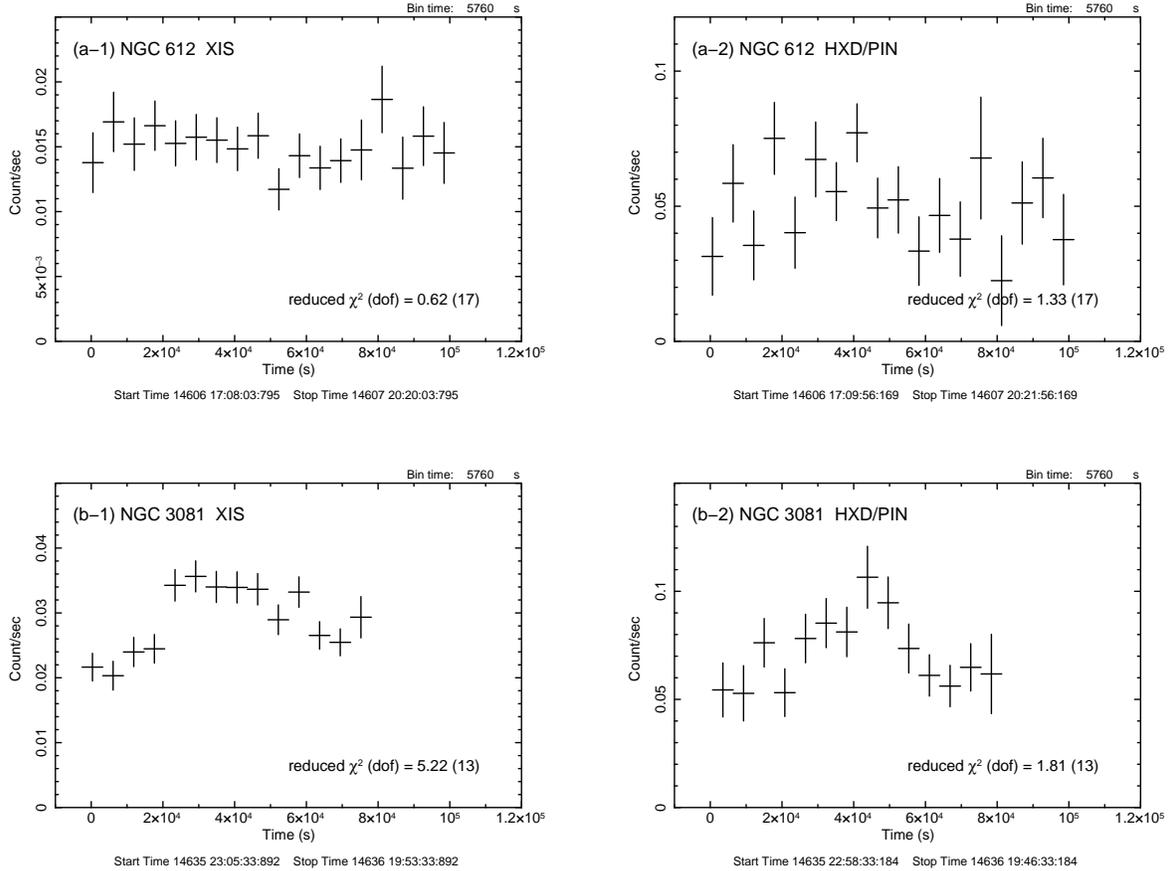}
\caption{
The background subtracted light curves of {\suzaku}.
One bin corresponds to 96 minutes. The numbers listed in each panel
represent the value of reduced $\chi^{2}$ with the degrees of freedom
for the constant flux hypothesis. \textit{Left}: The light curves of
the XIS in the 2--10 keV band. The data from the XIS-0 and XIS-3 are
summed. \textit{Right}: The light curves of the HXD/PIN in the 15--40 keV
band.\label{fig-lc}}
\end{figure*}

We analyze only the data of the XISs and the HXD/PIN, which covers the
energy band of 0.2--12 keV and 10--60 keV, respectively. The fluxes
above 50 keV are too faint to be detected with HXD/GSO.
Table~\ref{tab-observations} shows the log of the observations. The
net exposure of each target is about 45 ks.
Because XIS-2 became unoperatable on 2007 November 7
\citep{Dotani2007}, no XIS-2 data are available for both objects.
For the XIS observations, we applied spaced-row charge injection (SCI)
to improve the energy resolution \citep{Nakajima2008}; for instance, 
it reduces the full width at half maximum of the ${}^{55} \rm{Fe}$
calibration source from $\simeq 230 \ \rm{eV}$ 
to $\simeq 160 \ \rm{eV}$ for XIS-0 \citep{Ozawa2009}.
To constrain the broad band spectra above 60 keV, we also utilize the
{\swift}/BAT spectra covering the 15--200 keV band, integrated over
the first 22-months of {\swift} operations.

\subsection{Data Reduction}

The {\suzaku} data are analyzed by using \textit{HEAsoft} version 6.7
and the latest version of \textit{CALDB} on 2009 December 3. For the
XIS data, we analyze the version 2.2 cleaned events distributed by the
{\suzaku} pipeline processing team. In extraction of the light curves
and spectra, we set the source region as a circle around the detected
position with a radius of 1.5 arcmin, where about 75\% of the total source photons are
accumulated, to maximize the signal-to-noise ratio.
The background for the XIS data is taken from a source-free region in
the field of view with an approximately same offset angle from the
optical axis as the source.  For the non X-ray background of the HXD/PIN data,
we use the so-called ``tuned'' background model provided by the HXD
team. Its systematic errors are estimated to be $\simeq 0.97\%$ at a
$1 \sigma$ confidence level in the 15--40 keV band for a 40 ks
exposure \citep{Mizuno2008}. Since our exposures are $\approx 40 \
\rm{ks}$ or longer, we expect that the error is even smaller than this
value. The CXB spectrum simulated with the HXD/PIN response for a
uniformly extended emission is added to the non X-ray background 
spectrum.

\subsection{Light Curves}

Figure \ref{fig-lc} shows the background-subtracted light curves of
our targets obtained with the XIS and HXD/PIN in the 2--10 keV and
15--40 keV bands, respectively.  To minimize any systematic
uncertainties caused by the orbital change of satellite, we merge data
taken during one orbit ($\simeq 96$ minutes) into one bin. Then, to
check if there are any significant time variability during the
observations, we perform a simple $\chi^{2}$ test to each light curve
assuming a null hypothesis of a constant flux. The resultant reduced
$\chi^{2}$ value and the degrees of freedom are shown in each panel.
As noticed from Figure~\ref{fig-lc}, the 2--10 keV flux of NGC~3081
increased by a factor of 1.5 after $\simeq 20 \ \rm{ks}$ from the
start of the observation. Then, a flux decline is suggested between
$\simeq 50 \ \rm{ks}$ and $\simeq 60 \ \rm{ks}$ particularly in the
15--40 keV band.
Thus, we divide the observation of NGC~3081 into three different time
regions, 0--20 ks (Epoch 1), 20--60 ks (Epoch 2), and 60--80 ks (Epoch
3) measured from the observation start. By contrast, no significant
time variability on a time scale of hours are detected from
NGC~612. Hence, we analyze the time-averaged spectra over the whole
observation for NGC~612.

\subsection{BAT Spectra}

It is known that the incident photon spectra of Seyfert galaxies are
roughly approximated by a power law with an exponential cutoff (cutoff
power-law model), represented as $A E^{-\Gamma} \exp \left( -E /
E_{\rm{cut}} \right)$, where $A$, $\Gamma$, $E_{\rm{cut}}$ are the
normalization at 1 keV, photon index, and cutoff energy,
respectively. We analyze the {\swift}/BAT spectra in the 15--200 keV band to
constrain $E_{\rm{cut}}$. Here we take into account possible
contribution from a Compton reflection component from optically thick,
cold matter, utilizing the \textbf{pexrav} code
\citet{Magdziarz1995}. The relative intensity of the reflection
component to that of the intrinsic cutoff power-law component is
defined as $R \equiv \Omega / 2 \pi$, where $\Omega$ is the solid
angle of the reflector ($R = 1$ corresponds to the reflection from a
semi-infinite plane).

\begin{deluxetable}{ccc}
\tabletypesize{\scriptsize}
\tablecaption{Cutoff Energies ($E_{\rm{cut}}$) determined by the BAT spectra\label{tab-ecut}}
\tablewidth{0pt}
\tablehead{\colhead{$\Omega / 2 \pi$} & \colhead{NGC~612} & \colhead{NGC~3081}}
\startdata
$0$ & $> 315$ & $> 338$  \\
$\chi^{2} / \rm{d.o.f.}$ & $16.4 / 6$ & $11.6 / 6$ \\ \hline
$2$ & $> 319$ & $> 293$ \\
$\chi^{2} / \rm{d.o.f.}$ & $10.3 / 6$ & $5.8 / 6$ \\
\enddata
\tablecomments{The unit of $E_{\rm{cut}}$ is keV.}
\end{deluxetable}

In the analysis of the {\swift}/BAT spectra, we assume $R = 0$ or $2$
as the two extreme cases just to evaluate the effects of including the
reflection components, as done in Paper~I. The inclination angle is
fixed at $60^{\circ}$. To avoid strong coupling between the power-law
slope and cutoff energy, we fix the photon index at 1.9, the canonical
slope for AGNs \citep[e.g.,][]{Nandra1994}. Table~\ref{tab-ecut}
gives the fitting results for $E_{\rm{cut}}$; we find that
$E_{\rm{cut}}$ is greater than $\simeq 300 \ \rm{keV}$ for both
targets.\footnote{This conclusion is unchanged when
we fix the inclination angle at $30^{\circ}$ or $80^{\circ}$,
indicating that $E_{\rm{cut}}$ is not sensitive to the assumed
inclination angle.} Accordingly, we fix it at 300 keV (or 360 keV
for consistency with the Ikeda model) in the following spectral
analysis.

\section{Analytical Models}\label{sec-analytical-model}

We perform the spectral fitting to the {\suzaku} data in the same
manner as Paper I. We start with the simplest model for each target,
and if we find that the fit with a simple model does not give a
physically self-consistent picture or that the fit is significantly
improved by introducing additional parameters, then we adopt more
complicated models. We use only the {\suzaku} XIS and HXD/PIN data
throughout this stage, and finally perform the simultaneous fit of
XIS, HXD/PIN, and {\swift}/BAT spectra with the selected model to
obtain the best-fit parameters.

The spectra of FI-XISs are summed, and the relative normalization
between the FI-XISs and the PIN is fixed at 1.18 based on the
calibration of Crab Nebula \citep{Maeda2008}. Those of BI-XIS and BAT
against FI-XISs are set as free parameters. The Galactic absorption
($N_{\rm{H}}^{\rm{Gal}}$) is always included in the models, whose
hydrogen column density is fixed at values obtained from the H~I map
\citet{Kalberla2005}, available with the \textbf{nh} program in the
\textit{HEAsoft} package. We adopt the photoelectric absorption cross
section by \citet{Balucinska1992} (``bcmc''). Different from Paper~I,
we allow the iron abundance to be a free parameter by using
\textbf{zvphabs} because non Solar values (as defined
by \citealt{Anders1989}) are required to explain the {\suzaku} spectra,
while Solar abundances are adopted for the other metals throughout our
analysis.

\begin{figure}
\epsscale{1.0}
\plotone{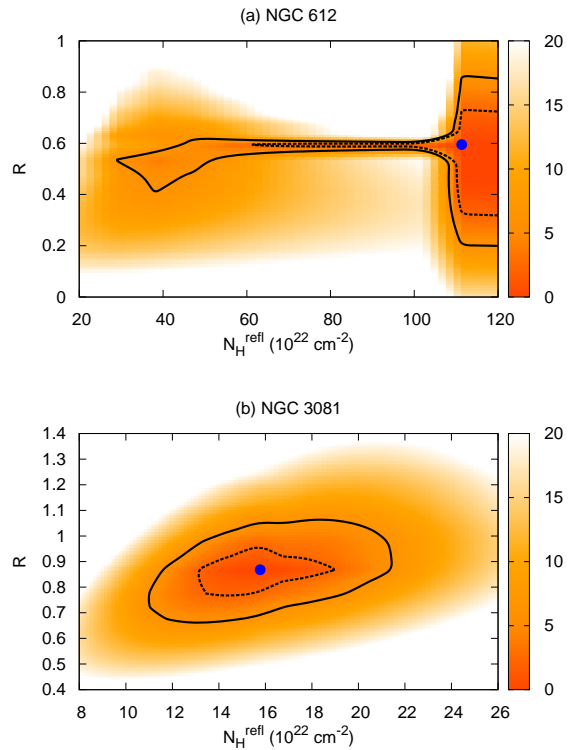}
\caption{
The confidence map in $\Delta \chi^{2}$ (color scale) with respect to
the strength of the reflection component ($R = \Omega / 2 \pi$) and
its absorption ($N_{\rm{H}}^{\rm{refl}}$) for
NGC~612 (top) and NGC~3081 (bottom). The dashed and solid curves
correspond to the 1$\sigma$ and 2$\sigma$ confidence
level for two interesting parameters, respectively.
\label{fig-R-nhrefl-contour}}
\end{figure}

We use the same three models as defined in Paper I, consisting of an
absorbed transmitted component, a scattered component, and/or 
an absorbed reflection component, with an iron-K emission line:
\begin{itemize}
 \item Model A: transmission + scattering + iron line,\footnote{In \textit{XSPEC} nomenclature, \textbf{zvphabs*zhighect*zpowerlw + const*zhighect*zpowerlw + zgauss}}
 \item Model B: transmission + scattering + iron line + absorbed reflection,\footnote{In \textit{XSPEC} nomenclature, \textbf{vzphabs*zhighect*zpowerlw
                + const*zhighect*powerlw + zgauss + zvphabs*pexrav}}
 \item Model C: transmission with dual absorber + scattering + iron line + absorbed reflection.\footnote{In \textit{XSPEC} nomenclature, \textbf{zvphabs*zpcfabs*zhighect*zpowerlw
                + const*zhighect*zpowerlw + zgauss + zvphabs*pexrav}}
\end{itemize}
In our analysis, we adopt an unabsorbed power law with the same
photon index as the incident continuum to describe the scattered
component, ignoring any emission lines from the photo-ionized gas.
Note that here we only introduce a single absorber for the reflection
component as the simplest approximation, although we expect both
absorbed and unabsorbed ones from the torus as well as that from the
accretion disk, as described in next Section.  The \textbf{pexrav}
component in each spectral model (see footnote) represents only the
reflection component not including the direct one by setting $R<0$,
and the inclination angle is fixed at 60$^\circ$.
Theoretically, the equivalent width ($\rm{E.W.}$) of the iron-K
emission line with respect to the reflection component, $\rm{E.W.}^{\rm{refl}}$, is expected to be $\sim$1 keV \citep{Matt1991}.
Since this value depends on the geometry of the reflector as well as 
the iron abundance, we regard the result as physically valid 
if $\rm{E.W.}^{\rm{refl}}$ = 0.5--2 keV.
No other emission lines than iron K$\alpha$ are significantly detected from the
spectra.  Considering calibration uncertainties in the energy
response of the XISs, we fix the $1 \sigma$ line width of the
iron-K$\alpha$ emission at the averaged value of the (apparent) line
width of the ${}^{55} \rm{Fe}$ calibration source at 5.9 keV: 45 eV
and 47 eV for NGC~612 and NGC~3081, respectively.

\subsection{NGC~612}

\begin{figure*}
\epsscale{1.0}
\plotone{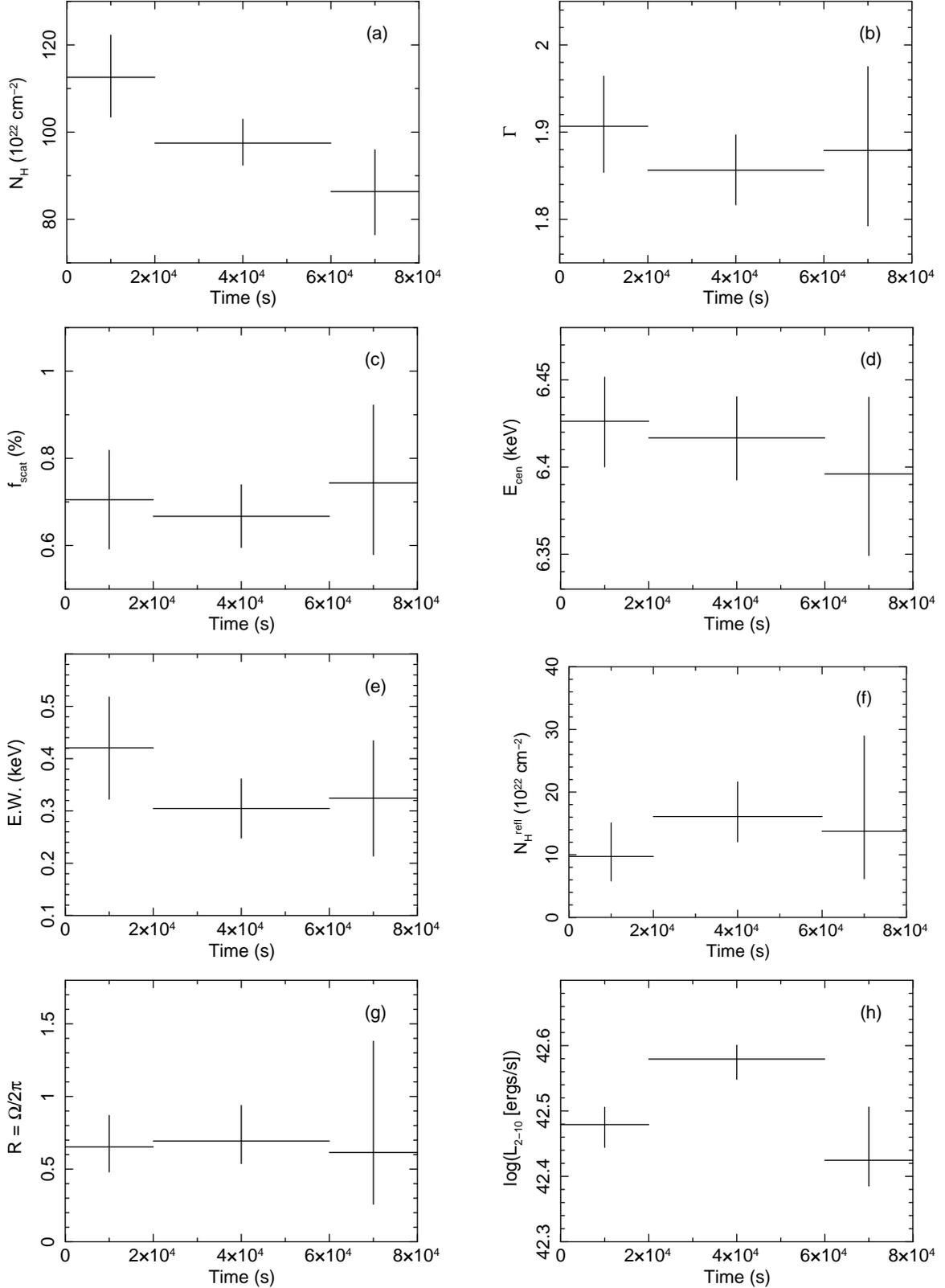}
\caption{
Time variability of the best-fit parameters of NGC~3081 obtained with
the analytical model: from left to right and top to bottom,
(a) the line-of-sight hydrogen column density for the transmitted component, 
(b) the power-law photon index,
(c) the fraction of the scattered component relative to the intrinsic power law,
(d) the center energy of the iron-K emission line at the rest frame,
(e) the equivalent width of the iron-K line with respect to the whole continuum,
(f) the line-of-sight hydrogen column density for the reflection component, 
(g) the solid angle of the reflection component,
(h) the 2--10 keV intrinsic luminosity corrected for the absorption.\label{fig-sw0959-var}}
\end{figure*}

Model~B is adopted as the most appropriate model of NGC~612. We obtain
$(\chi^{2}, \nu) = (90.8, 85)$ with Model~A and $(\chi^{2}, \nu) =
(86.9, 84)$ with Model B from the {\suzaku} spectra,
where $\nu$ is the degree of freedom. Thus, the improvement of the fit
by adding a reflection component is found to be significant at
$94\%$ confidence level by an F-test. No significant
improvement is found with Model C. For this target, the absorption to
the reflection component $N_{\rm{H}}^{\rm{refl}}$ is linked to that
for the transmitted component, because making them independent does
not give a better fit over the statistics (see below).
Since the E.W. of the iron-K line with respect to the
reflection component is $\rm{E.W.}^{\rm{refl}} = 0.7 \pm 0.1 \ \rm{keV}$, the
model is physically self-consistent; the iron abundance of NGC~612 is
roughly half of the Solar value, $0.54^{+0.10}_{-0.07}$, obtained from
the simultaneous fit of the {\suzaku} and BAT spectra.

To examine the degeneracy in the fitting parameters, in
Figure~\ref{fig-R-nhrefl-contour} (top) we show the confidence map
(in terms of $\Delta \chi^{2}$) with respect to the strength of the
reflection component ($R$) and its absorption
($N_{\rm{H}}^{\rm{refl}}$), based on the Model~B fit including the BAT
data. Here we do not link $N_{\rm{H}}^{\rm{refl}}$ to that of the
transmitted component ($N_{\rm{H}}$), and explore a region of
$N_{\rm{H}}^{\rm{refl}} < 1.2\times10^{24}$ cm$^{-2}$, the upper limit
obtained for $N_{\rm{H}}$.
The contours give the condidence levels at 1$\sigma$ and 2$\sigma$ for
two interesting parameters. As noticed, while a wide range of
$N_{\rm{H}}^{\rm{refl}} \ (\gtrsim 6 \times 10^{23} \ \rm{cm}^{-2},
1\sigma)$ is allowed, we can constrain the reflection strength to be
$R\simeq 0.6$ for $N_{\rm{H}}^{\rm{refl}} < 1.1
\times10^{24} \ \rm{cm}^{-2}$ and $R \simeq 0.3-0.7$ otherwise. The case
of an unabsorbed reflection component ($N_{\rm{H}}^{\rm{refl}} = 0$)
or no reflection component ($R = 0$) is rejected at $> 99\%$
confidence level, which corresponds to $\Delta \chi^{2} = 9.21$.

\subsection{NGC~3081}

\begin{figure*}
\epsscale{1.0}
\plotone{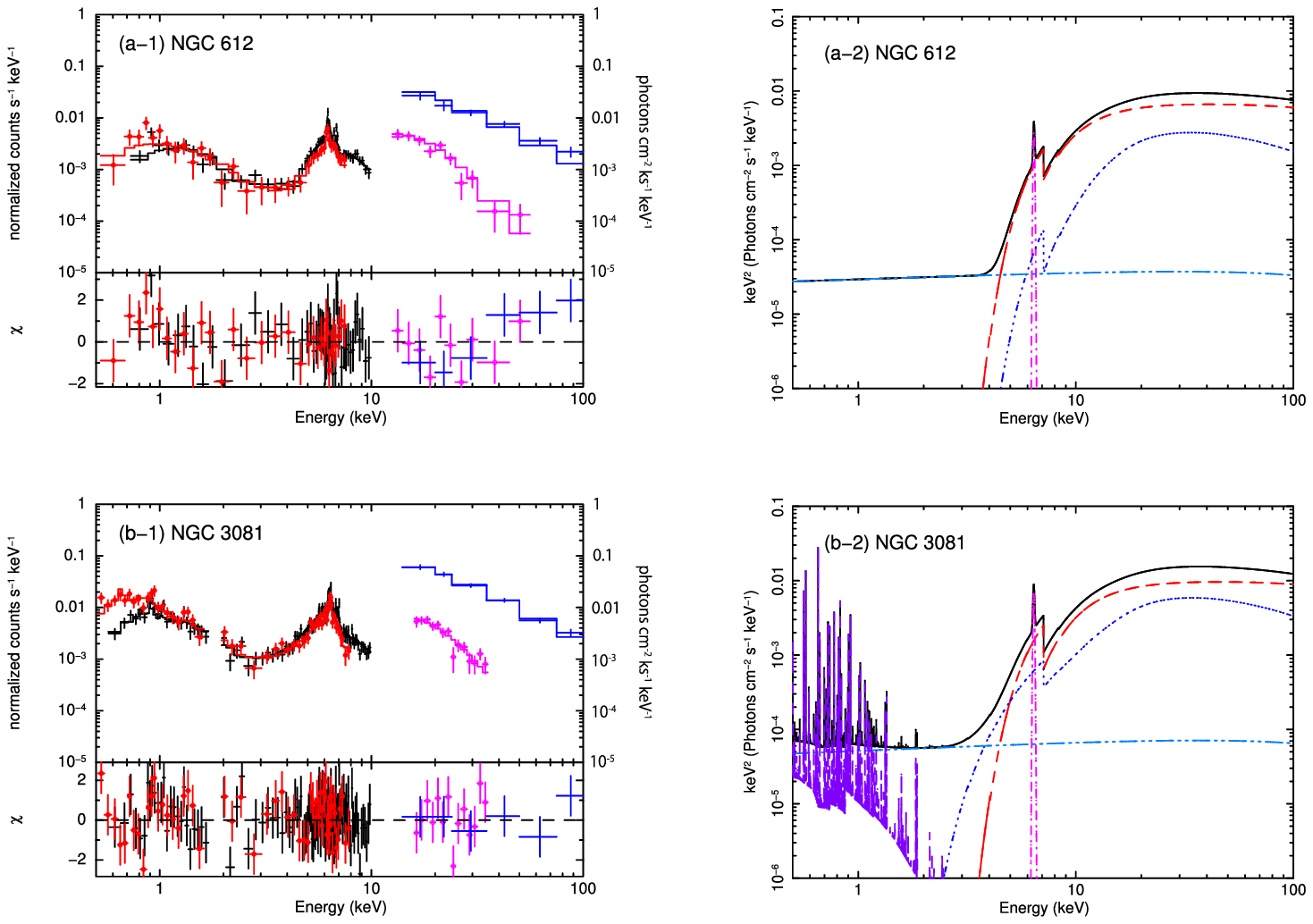}
\caption{
The observed spectra (left) and the best-fit spectra model (right) of
NGC~612 and NGC~3081. \textit{Left}: The black crosses, red filled circles,
magenta open circles, and blue crosses represent the data of the FI-XIS, BI-XIS, HXD/PIN,
and BAT, respectively, with 1$\sigma$ error bars. 
The spectra of the XIS and PIN are folded with the detector responses
in units of counts s$^{-1}$ keV$^{-1}$, while those of the BAT 
are {\it unfolded} spectra 
in units of photons cm$^{-2}$ ks$^{-1}$ keV$^{-1}$. The best-fit
models are plotted by solid curves, and the residuals in units of
$\chi$ are shown in the lower panels.  \textit{Right}: The best-fit
spectral model in units of $E F_{E}$ (where $E$ is the energy and
$F_{E}$ is the photon spectrum). The black, dashed red, dotted blue,
dot-dot-dashed cyan, dot-dashed magenta curves correspond to the
total, transmitted one, reflection component, scattered component, and
iron-K emission line, respectively. The purple dashed model below 2
keV in NGC~3081 represents the emission from an optically-thin thermal
plasma (see text). \label{fig-spectra-analytical}}
\end{figure*}

First, we analyze the {\suzaku} spectra integrated
over Epoch~1. We obtain $(\chi^{2}, \nu) = (120.1, 61)$ with Model A
and $(\chi^{2}, \nu) = (80.2, 59)$ with Model B, and thus the
improvement of $\chi^{2}$ is significant at $> 99\%$
confidence level by an F-test. No significant improvement is found
with Model C. Finally, since positive residuals remain in the energy
band below 1 keV, we add the
\textbf{vapec}\footnote{\texttt{http://cxc.harvard.edu/atomdb/}} in
\textit{XSPEC}, a spectral model from an optically-thin thermal
plasma, whose iron abundance is linked to that in the absorber of the
transmitted component. This yields a further significantly better fit
at $> 99\%$ confidence level with $(\chi^{2}, \nu) =
(50.7, 57)$. The E.W.\ of the iron-K line with respect to the
reflection component is $\rm{E.W.}^{\rm{refl}} = 1.5 \pm 0.3 \ \rm{keV}$,
which is self-consistent. Thus, we adopt Model B+\textbf{vapec} as the
best fit model for Epoch~1.

By fitting the Epoch~2 and 3 spectra of {\suzaku} with the same model,
we obtain acceptable fits with $(\chi^2, \nu) = (139.7, 114)$ and $(19.1,
30)$, respectively.
The best-fit parameters in the three epochs are plotted in
Figure~\ref{fig-sw0959-var} except for the iron abundance and those of
the thin thermal component, which are expected to show no time
variability.
We find only a weak indication that the column density
changed during the observation from $N_{\rm{H}} = (113 \pm 10) \times
10^{22} \ \rm{cm}^{-2}$ (Epoch~1), to $N_{\rm{H}} = (98 \pm 6) \times
10^{22} \ \rm{cm}^{-2}$ (Epoch~2), and then $N_{\rm{H}} = (86 \pm 10)\times
10^{22} \ \rm{cm}^{-2}$ (Epoch~3) in addition to the unabsorbed
power-law luminosity, which varied from $L_{\rm{2-10}} = (3.0 \pm 0.2)
\times 10^{42} \ \rm{ergs} \, \rm{s}^{-1}$ (Epoch~1), to $L_{\rm{2-10}} =
3.8^{+0.2}_{-0.3} \times 10^{42} \ \rm{ergs} \, \rm{s}^{-1}$ (Epoch~2),
and $L_{\rm{2-10}} = 2.7^{+0.5}_{-0.2} \times 10^{42} \ \rm{ergs} \,
\rm{s}^{-1}$ (Epoch~3) in the 2--10 keV band. The other parameters are
found to be consistent with being constant among the three epochs
within the errors.
The significance of the variability of the column density is marginal,
as the null hypothesis probability of a constant value is found to be
15\% from a $\chi^{2}$ test. Thus, we sum the Epoch 1, 2 and 3 data
of {\suzaku} and discuss the time averaged spectra in the following
analysis.

To best constrain the spectral parameters of NGC~3081, we perform the
simultaneous fit to the time-averaged {\suzaku} and BAT spectra with
the Model B+\textbf{vapec} model.  This yields $(\chi^{2}, \nu) =0
(199.28, 199)$ and $\rm{E.W.}^{\rm{refl}} = 1.0 \pm 0.1 \ \rm{keV}$, and
thus is physically self-consistent. The iron abundance with
respect to Solar is $0.89 \pm 0.07$, and the temperature
of the plasma is found to be $kT = 0.26 \pm 0.02 \ \rm{keV}$
with an emission measure of $n^{2} V \simeq 1.5 \times 10^{63} \
\rm{cm}^{-3}$. 

Figure~\ref{fig-R-nhrefl-contour} (bottom) shows the
confidence contour map with respect to $R$ and
$N_{\rm{H}}^{\rm{refl}}$ based on Model~B (including the BAT data) for
NGC~3081. Unlike the case of NGC~612, the solution
is well constrained ($R \simeq 0.8-1.0$, 1$\sigma$) and
we do not see strong degeneracy in the fitting parameters. Again,
neither the case of an unabsorbed reflection component
($N_{\rm{H}}^{\rm{refl}} = 0$), nor no reflection component ($R = 0$)
is allowed at $> 99\%$ confidence level.

\subsection{Results Summary of Analytical Models}

\begin{deluxetable}{cccc}
\tabletypesize{\scriptsize}
\tablecaption{Best-fit Spectral Parameters with Analytical Models\label{tab-parameters-analytical}}
\tablewidth{0pt}
\tablehead{\colhead{} & \colhead{} & \colhead{NGC~612} & \colhead{NGC~3081}}
\startdata
 & Best-fit model & B & B + vapec\tablenotemark{a} \\
(1) & $N_{\rm{H}}^{\rm{Gal}}$ ($10^{22} \ \rm{cm}^{-2}$) & $0.0195$ & $0.0388$ \\
(2) & $N_{\rm{H}}$ ($10^{22} \ \rm{cm}^{-2}$) & $111 \pm 5$ & $98 \pm 4$ \\
(3) & $Z_{\rm{Fe}}$ & $0.54^{+0.10}_{-0.07}$ & $0.89 \pm 0.07$ \\
(4) & $\Gamma$ & $1.90 \pm 0.04$ & $1.88 \pm 0.02$ \\
(5) & $f_{\rm{scat}}$ (\%) & $0.55 \pm 0.06$ & $0.73 \pm 0.07$ \\
(6) & $E_{\rm{cen}}$ (keV) & $6.42^{+0.03}_{-0.02}$ & $6.41 \pm 0.02$ \\
(7) & E.W. (keV) & $0.28 \pm 0.06$ & $0.35 \pm 0.05$ \\
(8) & $\rm{E.W.}^{\rm{refl}}$ (keV) & $0.7 \pm 0.1$ & $1.0 \pm 0.1$ \\
(9) & $N_{\rm{H}}^{\rm{refl}}$ ($10^{22} \ \rm{cm}^{-2}$) & $(= N_{\rm{H}})$ & $16^{+4}_{-3}$ \\
(10) & $R$ & $0.6 \pm 0.2$ & $0.9 \pm 0.2$ \\
(11) & $F_{\rm{2-10}}$ ($\rm{ergs} \, \rm{cm}^{-2} \, \rm{s}^{-1}$) & $1.6 \times 10^{-12}$ & $2.7 \times 10^{-12}$ \\
(12) & $F_{\rm{10-50}}$ ($\rm{ergs} \, \rm{cm}^{-2} \, \rm{s}^{-1}$) & $1.9 \times 10^{-11}$ & $3.1 \times 10^{-11}$ \\
(13) & $L_{\rm{2-10}}$ ($\rm{ergs} \, \rm{s}^{-1}$) & $3.0 \times 10^{43}$ & $3.0 \times 10^{42}$ \\
 & $\chi^{2} / \rm{d.o.f.}$ & $100.4 / 91$ & $199.3 / 199$ \\
\enddata
\tablenotetext{a}{An additional emission from an optically-thin thermal plasma is required,
modelled by the \textbf{vapec} code with a temperature of $kT = 0.26 \pm 0.02 \ \rm{keV}$ and an emission measure of
$1.5 \times 10^{63} \ \rm{cm}^{-3}$.
The iron abundance is linked to that in the absorber of the transmitted component (see text).}
\tablecomments{
(1) The hydrogen column density of Galactic absorption by \citet{Kalberla2005}.
(2) The line-of-sight hydrogen column density for the transmitted component.
(3) The iron abundance relative to the Solar value.
(4) The power-law photon index.
(5) The fraction of the scattered component relative to the intrinsic power law.
(6) The center energy of the iron-K emission line at the rest frame of the source redshift.
(7) The observed equivalent width of the iron-K line with respect to the whole continuum.
(8) The observed equivalent width of the iron-K line with respect to the reflection component
(9) The line-of-sight hydrogen column density for the reflection component.
(10) The relative strength of the reflection component to the transmitted one,
defined as $R \equiv \Omega / 2 \pi$, where $\Omega$ is the solid angle of the reflector viewed from the nucleus.
(11) The observed flux in the 2--10 keV band.
(12) The observed flux in the 10--50 keV band.
(13) The 2--10 keV intrinsic luminosity corrected for the absorption.
The errors are 90\% confidence limits for a single parameter.
}
\end{deluxetable}

We summarize the best-fit models and parameters in Table
\ref{tab-parameters-analytical}. The observed fluxes in the 2--10 keV and
10--50 keV bands, and the estimated 2--10 keV intrinsic luminosities
corrected for absorption are also listed. 
Figure~\ref{fig-spectra-analytical} (left) shows the observed spectra
of the FI-XIS (black), the BI-XIS (red), and the HXD/PIN (magenta)
folded with the detector response in units of $\rm{counts} \
\rm{s}^{-1} \ \rm{keV}^{-1}$, together with the {\it
unfolded} BAT spectra (blue) in units of $\rm{photons} \ \rm{cm}^{-2}
\ \rm{ks}^{-1} \ \rm{keV}^{-1}$. The best-fit models are superposed by
solid lines. In the lower panels, the corresponding data-to-model
residuals in units of $\chi$ (i.e., normalized by the $1 \sigma$
statistical error in each bin) are plotted.
Figure~\ref{fig-spectra-analytical} (right) 
shows the best-fit spectral models in units of $E F_{E}$ without
Galactic absorption, where the contribution of each component is
plotted separately; the black, red, blue, cyan, magenta curves
correspond to the total, transmitted component, reflection component,
scattered component, and iron-K emission line, respectively. For
NGC~3081, the additional soft component is also included in purple.

\begin{figure}[b]
\epsscale{1.0}
\plotone{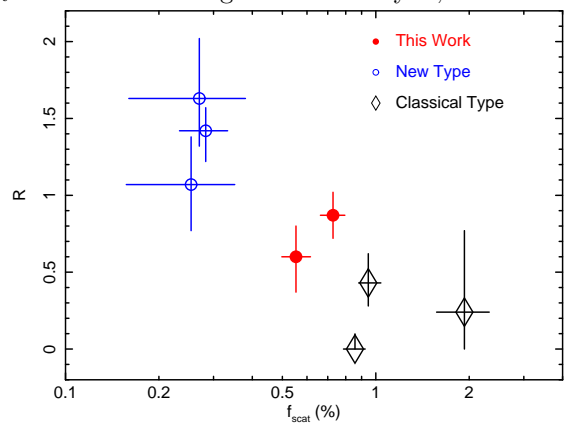}
\caption{
The correlation between the strength of the Compton reflection
component ($R = \Omega / 2 \pi$) and the fraction of the scattered
component ($f_{\rm{scat}}$) for our targets (filled circles), superposed
on the same figure taken from Paper I, where open circle and open
diamond represent ``new type'' ($R \gtrsim 0.8$ and $f_{\rm{scat}}
\lesssim 0.5\%$) and ``classical type'' ($R \lesssim 0.8$ and
$f_{\rm{scat}} \gtrsim 0.5\%$) AGNs, respectively.\label{fig-R-fscat}
}
\end{figure}

We find that the fraction of the scattered component with respect to
the transmitted one for both of NGC~612 and NGC~3081 are fairly
small (0.5\%--0.8\%). Also, they are heavily obscured with column
densities of $N_{\rm{H}} \simeq 10^{24} \ \rm{cm}^{-2}$, and hence can
be regarded as nearly ``Compton thick'' AGNs. Figure~\ref{fig-R-fscat}
shows the correlation between the reflection component $R$ and the
scattered component $f_{\rm{scat}}$, superposed on the same plot
presented in Paper~I. NGC~612 and NGC~3081 are located on a
similar position to each other in this plot. In Paper I, the authors
categorize the observed AGNs into two groups: ``new type'' with $R
\gtrsim 0.8$ and $f_{\rm{scat}} \lesssim 0.5\%$, and ``classical
type'' with $R \lesssim 0.8$ and $f_{\rm{scat}} \gtrsim 0.5\%$. Since
our targets are placed just between the two types, it is not clear to
which ``type'' these AGNs belong if the intrinsic distribution is
indeed distinct. It is also possible that the distribution is smooth
and they actually represent an intermediate class bridging the two types.

\begin{figure}[b]
\epsscale{1.0}
\plotone{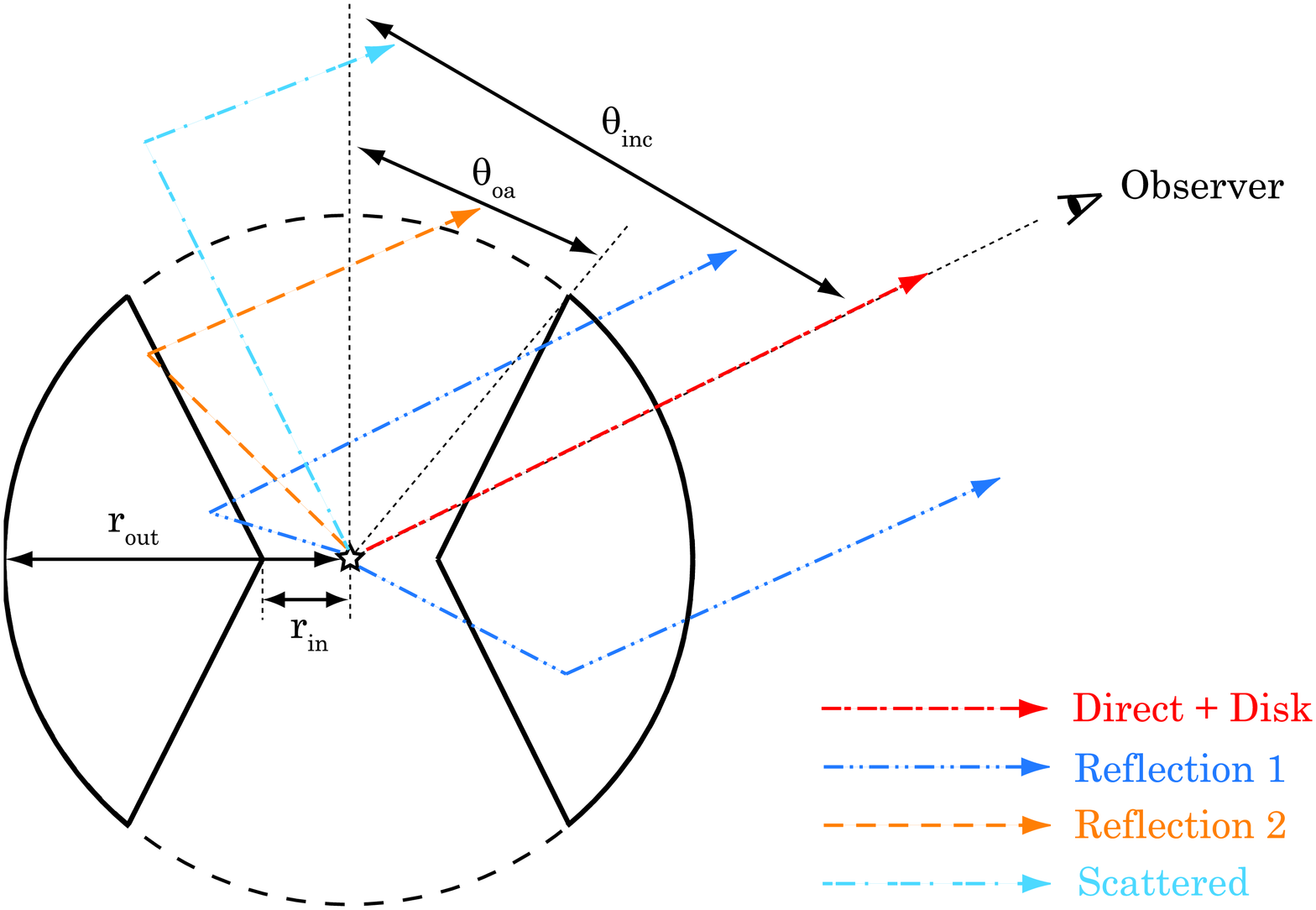}
\caption{
Cross-section view of the torus geometry assumed in \citet{Ikeda2009}.
The torus structure is characterized by the half-opening angle
$\theta_{\rm{oa}}$, the inclination angle of torus from an observer
$\theta_{\rm{inc}}$, the hydrogen column density viewed from the
equatorial plane $N_{\rm{H}}^{\rm{Eq}}$, and the ratio of
$r_{\rm{in}}$ to $r_{\rm{out}}$. The observed lights consist of a
transmitted component absorbed by the torus (dash-dashed red), a reflection
component from the accretion disk absorbed by the torus (dash-dashed red), two
reflection components by the torus; ``reflection component~1''
absorbed by the torus (dash-dot-dotted blue) and ``reflection component~2'' not
absorbed by it (dashed orange). We also consider a scattered component by the
surroinding gas, which is not absorbed by the torus (dash-dash-dotted cyan).
\label{fig-torus-geometry}}
\end{figure}

\section{Torus Model}\label{sec-torus-model}

\citet{Ikeda2009} performed a set of Monte Carlo simulations to calculate a
realistic reprocessed emission from the torus irradiated by a central
source, which is assumed to be a cutoff power-law model. In the
simulation, they assume a 3-dimensional axis-symmetric uniform torus as
illustrated in Figure \ref{fig-torus-geometry}. It is characterized by
the half-opening angle $\theta_{\rm{oa}}$, the inclination angle of
torus from an observer $\theta_{\rm{inc}}$, the hydrogen column
density viewed from the equatorial plane $N_{\rm{H}}^{\rm{Eq}}$, and 
the ratio $r$ of $r_{\rm{in}}$ to $r_{\rm{out}}$. 

\subsection{Method}

To perform spectral fitting on XSPEC with the \citet{Ikeda2009} model,
we utilize an \textbf{atable} model where the resultant spectra from
Monte Carlo calculation are stored at grids of the torus parameters in
FITS files. Here we refer to the results of the continuum reprocessed from
the torus (``torus-reflection component'') and those of iron-K
fluorescence line, assuming the geometry with $r \equiv r_{\rm{in}} /
r_{\rm{out}} = 0.01$.
The cutoff energy is fixed at $E_{\rm{cut}} = 360 \ \rm{keV}$
throughout our analysis, which is consistent with the contraints from the
{\swift}/BAT spectra for both targets ($> 300$ keV). Since the fit is
performed only in the 0.5--100 keV range where the table model is
available, the choice of $E_{\rm{cut}}$ hardly affects our
results as far as it is higher than $\sim$300 keV.
We consider two cases for the elemental abundances; ``Solar''
abundances (as defined by \citealt{Anders1989}) and ``Subsolar'' ones
where only those of iron and nickel are set to be 0.5 times Solar
values. Finally, the table files have five free parameters;
$N_{\rm{H}}^{\rm{Eq}}$, $\theta_{\rm{oa}}$, $\theta_{\rm{inc}}$, the
photon index $\Gamma$ of the incident continuum, and its normalization at
1 keV.

In the assumed geometry of the torus, the line-of-sight hydrogen column density
$N_{\rm{H}}$ for the transmitted component is related to that along
the equatorial plane ($N_{\rm{H}}^{\rm{Eq}}$) via equation (3) in
\citet{Ikeda2009}:
\begin{equation}
 \frac{N_{\rm{H}}}{N_{\rm{H}}^{\rm{Eq}}} = \frac{r \left( \cos \theta_{\rm{inc}} - \cos \theta_{\rm{oa}} \right) + \sin \left( \theta_{\rm{inc}} - \theta_{\rm_{oa}} \right)}{\left( 1 - r \right) \left\{ r \cos \theta_{\rm{inc}} + \sin \left( \theta_{\rm{inc}} - \theta_{\rm{oa}} \right) \right\}} \label{eq-nh}.
\end{equation}
We introduce \textbf{torusabs} (for the fixed Solar abundances) and
\textbf{vtorusabs} models (for variable abundances) as local models of
\textit{XSPEC} to represent photoelectric absorption of the
transmitted component, whose line-of-sight 
column density is related to the torus
parameters according to the above equation. 
In these models we take into account Compton scattering processes in
addition to photoelectric absorption, and hence they can be reliably
used even for the Compton thick case. We adopt the photoelectric
absorption cross section by \citet{Verner1996} for consistency with
\citet{Ikeda2009}.
As noted by \citet{Ikeda2009}, the cross section by \citet{Verner1996}
is more accurate for energies above 10 keV than that by
\citet{Balucinska1992}, and is nearly equal to the NIST XCOM
database.\footnote{\texttt{http://www.nist.gov/pml/data/xcom/index.cfm}}
Since the results become physically meaningless for
obscured AGNs if we obtain $\theta_{\rm{inc}} < \theta_{\rm{oa}}$, we
impose the condition that $\theta_{\rm{inc}} \geq \theta_{\rm{oa}} +
1^{\circ}$ in the fitting process.

The fraction of the scattered component to the transmitted component,
$f_{\rm{scat}}$, should be proportional to the opening solid angle of
the torus if the column density of the scattering gas is constant.
Thus, we have the constraint that
\begin{equation}
\cos \theta_{\rm{oa}} = 1 - \frac{f_{\rm{scat}}}{f_{\rm{scat, 0}}} \left( 1 - \cos \theta_{\rm{oa, 0}} \right), \label{eq-theta-oa}
\end{equation}
where we normalize $f_{\rm{scat}} = f_{\rm{scat, 0}}$ at
$\theta_{\rm{oa}} = \theta_{\rm{oa, 0}} \equiv 45^{\circ}$.
Similarly, we also developed the \textbf{fscat} model for \textit{XSPEC}
to calculate the normalization of the scattered emission as a function
of two parameters, $\theta_{\rm{oa}}$ and $f_{\rm{scat, 0}}$.
Here the normalization parameter $f_{\rm{scat, 0}}$ reflects the
averaged column density of the scattering gas and is treated 
as a free parameter. We allow it to vary within 0.1\%--5\%; note that 
a typical value in Seyfert 2 galaxies is $f_{\rm{scat}} = 3\%$
\citep{Guainazzi2005}.

In addition to the reprocessed emission from the torus, we should also
expect a reflection component from the accretion disk in AGN
spectra. Thus, in the transmitted component, we include this effect by
utilizing the \textbf{pexrav} model \citep{Magdziarz1995}, which is
appropriate to represent the reflection component from semi-infinite
plane, like that from accretion disks.  Here we fix the strength of
the disk reflection to be $R \equiv \Omega / 2 \pi = 1$, where
$\Omega$ is the solid angle of the accretion disk.  The inclination
angle of the accretion disk is linked to that of the torus. Although
the contribution of this component is not included in the incident
photon spectrum in the \citet{Ikeda2009} model, the effects on the ``torus
reflection'' spectra are only second-order and are negligible.
For simplicity, hereafter we refer to the total spectrum including
the reflection component from the accretion disk as ``torus model''.

\begin{figure*}
\epsscale{1.0}
\plotone{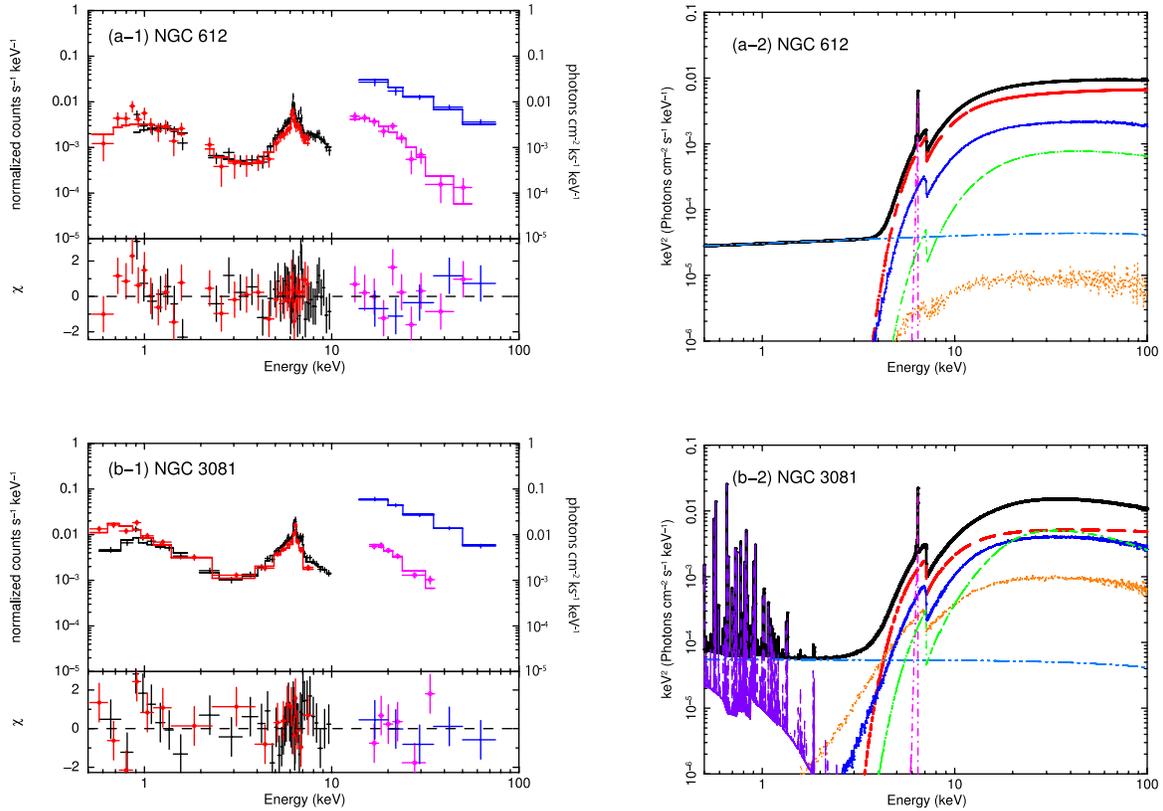}
\caption{
The observed spectra (left) and the best-fit spectral model (right)
with the torus model. \textit{Left}: same as
Figure~\ref{fig-spectra-analytical} (left). \textit{Right}: the best-fit
spectral model in units of $E F_{E}$ (where $E$ is the energy and
$F_{E}$ is the photon spectrum); total (thick black), transmitted component
(thick dashed red), reflection component from the accretion disk (thin dot-dashed
green), torus reflection component~1 (thin blue), torus reflection
component~2 (thin dotted orange), scattered component (thin dot-dot-dashed
cyan), iron-K emission line (thin dot-dashed magenta).  The purple dashed
curve below 2 keV in NGC~3081 represents the emission from an optically-thin
thermal plasma.}
\label{fig-spectra-torus-const}
\end{figure*}

To summarize, we can write the torus model of the photon spectrum $F \left( E \right)$ without the Galactic absorption as follows:
\begin{eqnarray}
 &F \left( E \right) = \nonumber \\
                    & \exp \left\{ - N_{\rm{H}} \left( N_{\rm{H}}^{\rm{Eq}}, \theta_{\rm{oa}}, \theta_{\rm{inc}} \right) \sigma \left( E \right) \right\} I \left( E \right) \nonumber \\
                    &+ f_{\rm{scat}} \left( \theta_{\rm{oa}}, f_{\rm{scat, 0}} \right) I \left( E \right) \nonumber \\
                    &+ \exp \left\{ - N_{\rm{H}} \left( N_{\rm{H}}^{\rm{Eq}}, \theta_{\rm{oa}}, \theta_{\rm{inc}} \right) \sigma \left( E \right) \right\} R_{\rm{disk}} \left( \theta_{\rm{inc}}, E \right) \nonumber \\
                    &+ R_{\rm{refl, 1}} \left( N_{\rm{H}}^{\rm{Eq}}, \theta_{\rm{oa}}, \theta_{\rm{inc}} \right) \nonumber \\
                    &+ R_{\rm{refl, 2}} \left( N_{\rm{H}}^{\rm{Eq}}, \theta_{\rm{oa}}, \theta_{\rm{inc}} \right) \nonumber \\
                    &+ \varepsilon_{\rm{Fe}} L_{\rm{Fe}} \left( N_{\rm{H}}^{\rm{Eq}}, \theta_{\rm{oa}}, \theta_{\rm{inc}} \right) \nonumber \\
                    &+ S \left( E \right), \label{eq-spectra-torus}
\end{eqnarray}
where  $I \left( E \right) \equiv A E^{- \Gamma} \exp \left( -E/E_{\rm{cut}}
\right)$ is the intrinsic cutoff power-law component,
$A$ is the normalization parameter of the intrinsic cutoff power
law at 1 keV, $N_{\rm{H}}^{\rm{Eq}}$ is the hydrogen column density of
the torus viewed from the equatorial plane, $\theta_{\rm{oa}}$ is the
half-opening angle of the torus, $\theta_{\rm{inc}}$ is the
inclination angle of the torus, $f_{\rm{scat, 0}}$ is the normalization
parameter at $\theta_{\rm{oa}} = 45^{\circ}$, $N_{\rm{H}} \left(
N_{\rm{H}}^{\rm{Eq}}, \theta_{\rm{oa}}, \theta_{\rm{inc}} \right)$ is
the absorption column density for the transmitted component, $\sigma
\left( E \right)$ is the cross section of photoelectric absorption,
$f_{\rm{scat}} \left( \theta_{\rm{oa}}, f_{\rm{scat, 0}} \right)$ is the scattered
fraction, $R_{\rm{disk}} \left( \theta_{\rm{inc}}, E \right)$ is the
Compton reflection component from the accretion disk, 
$R_{\rm{refl, 1}} \left( N_{\rm{H}}^{\rm{Eq}}, \theta_{\rm{oa}},
\theta_{\rm{inc}} \right)$ is the torus-reflection component~1 (see
Figure \ref{fig-torus-geometry}), $R_{\rm{refl, 2}} \left(
N_{\rm{H}}^{\rm{Eq}}, \theta_{\rm{oa}}, \theta_{\rm{inc}} \right)$ is
the torus-reflection component~2 (see Figure
\ref{fig-torus-geometry}), $\varepsilon_{\rm{Fe}}$ is the
normalization parameter for the iron-K emission line, $L_{\rm{Fe}}
\left( N_{\rm{H}}^{\rm{Eq}}, \theta_{\rm{oa}}, \theta_{\rm{inc}}
\right)$ is the iron-K emission line, and $S \left( E \right)$
represents additional soft components (for the case of NGC~3081
where the \textbf{apec} model is used).
We allow the relative normalization of the iron-K line
$\varepsilon_{\rm{Fe}}$ with respect to the torus-reflection
components to float between 0.5 and 2, in order to cover the iron
abundance range over the fixed values (0.5 or 1.0) and to model the effects of time
variability between the transmitted component and the averaged
reprocessed emission.

In short, Equation (\ref{eq-spectra-torus}) is expressed as
absorbed transmission\footnote{In \textit{XSPEC} nomenclature, \textbf{torusabs*zhighect*zpowerlw}} +
scattering\footnote{In \textit{XSPEC} nomenclature, \textbf{fscat*zhighect*zpowerlw}} +
absorbed accretion disk reflection\footnote{In \textit{XSPEC} nomenclature, \textbf{torusabs*toruspexrav}} +
absorbed torus reflection\footnote{In \textit{XSPEC} nomenclature, \textbf{atable\{refl1\_torus.fits\}}} +
unabsorbed torus reflection\footnote{In \textit{XSPEC} nomenclature, \textbf{atable\{refl2\_torus.fits\}}} +
iron line\footnote{In \textit{XSPEC} nomenclature, \textbf{constant*atable\{refl\_fe\_torus.fits\}}}.
There are eight free parameters: $\theta_{\rm{inc}}$,
$\theta_{\rm{oa}}$, $N_{\rm{H}}^{\rm{Eq}}$, $f_{\rm{scat, 0}}$,
$E_{\rm{cen}}$, $\varepsilon_{\rm{Fe}}$, $A$, and $\Gamma$.
In the spectral fit, we employ the \textit{MINUIT MIGRAD} method
(``migrad'') as the fitting algorithm, which is found to be more
stable than the standard \textit{Levennerg-Marquardt} method
(``leven'') in our case utilizing the numerical models.
Both of the {\suzaku} and BAT spectra (but below 100
keV) are used throughout this section.

\subsection{Application}

\subsubsection{NGC~612}

Figure~\ref{fig-spectra-torus-const} plots the best-fit torus model,
whose parameters are summarized in
Table~\ref{tab-parameters-torus-const}. The torus model with
``Subsolar'' abundances is adopted for this target, based on the
fitting result with the analytical models (Section~3). We find that
$\theta_{\rm{oa}} \simeq 60^{\circ}-70^{\circ}$ and $\theta_{\rm{inc}} \gtrsim
76^{\circ}$. As shown in Figure~\ref{fig-spectra-torus-const},
the contribution from the reflection component~2 (unabsorbed one) is very small,
suggesting that we are seeing the target from
an edge-on angle. In such case, the inclination angle
$\theta_{\rm{inc}}$ can be poorly determined above a certain
threshold, because the observed line-of-sight hydrogen column density
is rather insensitive to $\theta_{\rm{inc}}$ for a given
$N_{\rm{H}}^{\rm{Eq}}$ value. Actually, we find that the column density
in the equatorial plane, $N_{\rm{H}}^{\rm{Eq}} \simeq 10^{24.1} \
\rm{cm}^{-2}$, is close to that along the line of sight ($N_{\rm{H}}$)
as estimated from the analytical model fit.
We also find that $f_{\rm{scat, 0}} \simeq 0.14\%$, which indicates
that the amount of scattering gas around the nucleus is remarkably
small.

\subsubsection{NGC~3081}

The best-fit parameters with the torus model for NGC~3081 are
summarized in Table~\ref{tab-parameters-torus-const} and the model is
plotted in~Figure \ref{fig-spectra-torus-const}. For this target, we
adopt the ``Solar'' abundance tables based on the analytical model fit. We
obtain $\theta_{\rm{oa}} \simeq 15^{\circ}$, $\theta_{\rm{inc}} \simeq
19^{\circ}$, $N_{\rm{H}}^{\rm{Eq}} \simeq 10^{24.0} \
\rm{cm}^{-2}$, and $f_{\rm{scat, 0}} \simeq 4.6\%$.
By contrast to NGC~612, $f_{\rm{scat, 0}}$ is rather
large, and there is a significant contribution from the unabsorbed
reflection component, indicating that
the inclination angle must be close to the torus opening angle. The
results are consistent with the picture proposed for the ``new type''
AGN SWIFT J0601.9--8636 (ESO 005--G004) by \citet{Ueda2007}, that the
nucleus is deeply buried in the geometrically thick torus and is
observed from a rather face-on angle.

\begin{deluxetable}{cccc}
\tabletypesize{\scriptsize}
\tablecaption{Best-fit Spectral Parameters with Torus Model\label{tab-parameters-torus-const}}
\tablewidth{0pt}
\tablehead{\colhead{} & \colhead{} & \colhead{NGC~612} & \colhead{NGC~3081}}
\startdata
(1) & Table model & Subsolar & Solar + apec\tablenotemark{a} \\
(2) & $N_{\rm{H}}^{\rm{Gal}}$ ($10^{22} \ \rm{cm}^{-2}$) & $0.0195$ & $0.0388$ \\
(3) & $N_{\rm{H}}^{\rm{Eq}}$ ($10^{22} \ \rm{cm}^{-2}$) & $113^{+10}_{-8}$ & $91^{+10}_{-9}$ \\
(4) & $\theta_{\rm{oa}}$\tablenotemark{b} (degrees) & $70 \ (> 58)$ & $15 \pm 2$ \\
(5) & $\theta_{\rm{inc}}$ (degrees) & $87 \ (> 76)$ & $19^{+7}_{-1}$ \\
(6) & $\Gamma$ & $1.9 \pm 0.1$ & $2.0 \pm 0.1$ \\
(7) & $f_{\rm{scat, 0}}$ (\%) & $0.14 \ (< 0.19)$ & $4.6 \ (> 3.1)$ \\
(8) & $E_{\rm{cen}}$ (keV) & $6.43^{+0.11}_{-0.02}$ & $6.42^{+0.03}_{-0.04}$ \\
(9) & $\varepsilon_{\rm{Fe}}$ & $1.6^{+0.4}_{-0.6}$ & $0.57 \ (< 0.69)$ \\
 & $\chi^{2} / \rm{d.o.f.}$ & $84.3 / 83$ & $209.0 / 200$ \\
\enddata
\tablenotetext{a}{An additional emission from an optically-thin thermal plasma with Solar abundances is required, modelled by the \textbf{apec} code
with a temperature of $kT = 0.25 \pm 0.02 \ \rm{keV}$ and an emission measure of $1.5 \times 10^{63} \ \rm{cm}^{-3}$ (see text).}
\tablenotetext{b}{The range of the $\theta_{\rm{oa}}$ is limited to $< 70^{\circ}$ in the torus model.}
\tablecomments{
(1) The table model used in the fit. ``Solar'' means the table with Solar abundances, while ``Subsolar'' means the table with 0.5-time iron and nickel abundances
with respect to Solar ones.
(2) The hydrogen column density of Galactic absorption by \citet{Kalberla2005}.
(3) The hydrogen column density of the torus viewed from the equatorial direction.
(4) The half opening angle of the torus.
(5) The inclination angle of the torus.
(6) The power-law photon index.
(7) The fraction of the scattered component relative to the intrinsic power law when the half opening angle of the torus is $45^{\circ}$.
(8) The center energy of the iron-K emission line at the rest frame of the source redshift.
(9) The relative strength of the iron-K emission line to that predicted by the torus model.
The errors are 90\% confidence limits for a single parameter.
}
\end{deluxetable}

\section{Discussion and Conclusion}\label{sec-discussion}

With {\suzaku} follow-up, we have obtained the best-quality broad band
spectra covering the 0.5--60 keV band of two {\swift}/BAT AGNs,
NGC~612 and NGC~3081. First, we found a range in  the iron abundance; NGC~612 has about 0.5 times Solar abundance of iron
(where ``Solar'' corresponds to Fe/H = $4.68\times10^{-5}$), which is
significantly smaller than that of NGC~3081. Applying the analytical
models, we find that these objects are nearly Compton thick AGNs with
$N_{\rm{H}} \simeq 10^{24} \ \rm{cm}^{-2}$ and the fraction of the
scattered component with respect to the transmitted component is
small, $f_{\rm{scat}} < 0.8\%$, suggesting that these belong to
``hidden'' population according to \citet{Winter2009}. Plotting the
results in the $f_{\rm{scat}}$ versus reflection strength ($R$) plane,
we find these two targets are located just between those occupied by
``new type'' (geometrically thick tori) and ``classical type'' AGNs
defined in Paper~I, implying that they would be an intermediate class
bridging the two types.
We need a larger sample to reveal the true distribution of the
whole AGN population in this plane. In this
context, simultaneous broad band observations of more ``new type''
candidates are important to examine their reflection strengths, such
as those with small scattering fractions identified from the {\xmm}
catalog \citep{Noguchi2009}.

To further investigate the details of the torus geometry of the two
AGNs, we apply numerical spectral models based on Monte Carlo
simulation where a simple 3-dimensional geometry of the torus is
assumed, following the work by \citet{Ikeda2009} and
\citet{Awaki2009}. We also consider the Compton reflection component
from the accretion disk. To our knowledge, this is the first time all
effects both from the torus and disk are self-consistently considered
in spectral analysis of obscured AGNs. It is remarkable that we are
able to reproduce the observed spectra quite well with this torus
model, which has only 3 free geometrical parameters; the opening
angle, inclination, and equatorial column density.

The column density along the equator plane is found to be $N_{\rm
H}\approx 10^{24} \ \rm{cm}^{-2}$ for both sources, which is also
similar to that found from the Seyfert 2 galaxy Mrk~3 by
\citet{Ikeda2009}. The relative absence of higher column densities, though very limited in number, may
be consistent with the fact that even hard X-rays above 10 keV have a
bias against detecting heavily Compton thick AGNs with $N_{\rm{H}}
\sim 10^{25} \ \rm{cm}^{-2}$, unless the sample is limited to the very
local universe \citep{Malizia2009}. Thus, a majority of {\swift}/BAT
AGNs do not have extremely Compton thick tori defined at the equator
plane unless observed from a face-on angle. Future sensitive hard
X-ray surveys may start to pick up such populations, whose number
density and cosmological evolution are still open questions.

The analysis with the torus model suggest that the torus geometry of
the two targets may be different in spite of the very similar results
obtained from the analytical models. Our results confirm that the
fundamental assumption of the unified model where the opening angles
are all the same is too simple. For NGC~612, we find that the opening
angle is relatively large ($\simeq 60^{\circ}-70^{\circ}$) and the
object is observed from an edge-on angle, consistent with a picture of
``classical type'' Seyfert 2 galaxies. Similar torus parameters are
obtained for Mrk~3 by \citet{Ikeda2009}. By contrast, the torus
opening angle of NGC~3081 is much smaller ($\simeq 15^\circ$), and we
observe it from a face-on angle. This implies that NGC~3081 is closer
to a ``new type'' AGN discovered by \citet{Ueda2007} surrounded by a
geometrically thick torus. This picture for NGC~3081 is consistent
with the time variability of the column density, because we are seeing
the thinnest part of the torus that is expected to be highly patchy
\citep{Risaliti2002}.
We note, however, that the best-fit torus parameters
we obtain from the present analysis should not be taken at their face
values, which could depend on the initial assumption of the torus
geometry. For instance, as discussed in \citet{Ikeda2009}, if we
assume a lower value for $r (\equiv r_{\rm{in}} / r_{\rm{out}})$ than
0.01, we would obtain a slightly larger half-opening angle
$\theta_{\rm{oa}}$ for the same $N_{\rm{H}}^{\rm{Eq}}$ and
$\theta_{\rm{inc}}$ to account for the increased contribution of the
unabsorbed reflection component.

Since the observed scattering fraction is similar between the two
targets, this difference in the torus opening angle indicates that the
amount of scattering gas around the nucleus is much smaller in NGC~621
than in NGC~3081, as represented in the obtained $f_{\rm{scat, 0}}$
value, $f_{\rm{scat, 0}} < 0.2\%$ for NGC~612 and $f_{\rm{scat, 0}} >
3\%$ for NGC~3081. The small amount of the gas in NGC~612 may be
consistent with its classification as a ``weak emission line'' radio
galaxy, where the jets expel the surrounding gas. By contrast, the
detection of the optically-thin components in NGC~3081 could represent
the abundance of the ambient gas around the nucleus.

An important implication from the present study is that the
classification of different types of tori (e.g., geometrically thin or
thick) based solely on the scattered fraction may be difficult in some
cases. Our work has demonstrated the power of the application of
numerical torus models based on Monte Carlo simulation to best extract
the physical view of the nucleus beyond the simple phenomenological
spectral analysis, although caution must be paid because we have
considered only the simplest geometry by assuming a uniform density.
Combinations of the high quality broad band X-ray spectra with more
realistic numerical simulations will be a key approach for further
understanding of the nature of AGNs.

\acknowledgments

This work was partly supported by the Grant-in-Aid for JSPS Fellows
for young researchers (SE), Scientific Research 20540230 (YU),
21244017 (HA), and 20740109 (YT), and by the grant-in-aid for the
Global COE Program ``The Next Generation of Physics, Spun from
Universality and Emergence'' from the Ministry of Education, Culture,
Sports, Science and Technology (MEXT) of Japan.



\begin{thebibliography}{}
\bibitem[Anders \& Grevesse(1989)]{Anders1989} Anders, E., \& Grevesse, N.\ 1989, \gca, 53, 197
\bibitem[Awaki et al.(2009)]{Awaki2009} Awaki, H., Terashima, Y., Higaki, Y., \& Fukazawa, Y.\ 2009, \pasj, 61, 317
\bibitem[Balucinska-Church \& McCammon(1992)]{Balucinska1992} Balucinska-Church, M. \& McCammon, D. 1992, \apj, 400, 699
\bibitem[Bassani et al.(2006)]{Bassani2006} Bassani, L., et al.\ 2006, \apjl, 636, L65
\bibitem[Comastri et al.(2009)]{Comastri2009} Comastri, A., Iwasawa, K., Gilli, R., Vignali, C., \& Ranalli, P.\ 2009, arXiv:0910.1025
\bibitem[Comastri et al.(2010)]{Comastri2010} Comastri, A., Iwasawa, K., Gilli, R., Vignali, C., Ranalli, P., Matt, G., \& Fiore, F.\ 2010, \apj, 717, 787
\bibitem[Dotani et al.(2007)]{Dotani2007} Dotani, T. \& the XIS team 2007, JX-ISAS-SUZAKU-MEMO-2007-08
\bibitem[Eguchi et al.(2009)]{Eguchi2009} Eguchi, S., Ueda, Y., Terashima, Y., Mushotzky, R., \& Tueller, J.\ 2009, \apj, 696, 1657
\bibitem[Fanaroff \& Riley(1974)]{Fanaroff1974} Fanaroff, B.~L., \& Riley, J.~M.\ 1974, \mnras, 167, 31P
\bibitem[Freeman et al.(2000)]{Freeman2000} Freeman, T., Byrd, G., \& Ousley, D.\ 2000, IAU Colloq.~174: Small Galaxy Groups, 209, 325
\bibitem[Gilli et al.(2007)]{Gilli2007} Gilli, R., Comastri, A., \& Hasinger, G.\ 2007, \aap, 463, 79
\bibitem[Gopal-Krishna \& Wiita(2000)]{Gopal2000} Gopal-Krishna, \& Wiita, P.~J.\ 2000, \aap, 363, 507
\bibitem[Guainazzi et al.(2005)]{Guainazzi2005} Guainazzi, M., Matt, G., \& Perola, G.~C.\ 2005, \aap, 444, 119
\bibitem[Hopkins et al.(2005)]{Hopkins2005} Hopkins, P.~F., Hernquist, L., Cox, T.~J., Di Matteo, T., Martini, P., Robertson, B., \& Springel, V.\ 2005, \apj, 630, 705
\bibitem[Ikeda et al.(2009)]{Ikeda2009} Ikeda, S., Awaki, H., \& Terashima, Y.\ 2009, \apj, 692, 608
\bibitem[Joshi et al.(1989)]{Joshi1989} Joshi, U.~C., Jain, R., \& Deshpande, M.~R.\ 1989, Active Galactic Nuclei, 134, 321
\bibitem[Kalberla et al.(2005)]{Kalberla2005} Kalberla, P.~M.~W., Burton, W.~B., Hartmann, D., Arnal, E.~M., Bajaja, E., Morras, R., P\"{o}ppel, W.~G.~L.\ 2005, \aap, 440, 775
\bibitem[Krivonos et al.(2007)]{Krivonos2007} Krivonos, R., Revnivtsev, M., Lutovinov, A., Sazonov, S., Churazov, E., \& Sunyaev, R.\ 2007, \aap, 475, 775
\bibitem[Maeda et al.(2008)]{Maeda2008} Maeda, Y., Someya, K., Ishida, M., \& the XRT team, Hayashida, K., Mori, H., \& the XIS team 2008, JX-ISAS-SUZAKU-MEMO-2008-06
\bibitem[Magdziarz \& Zdziarski(1995)]{Magdziarz1995} Magdziarz, P., \& Zdziarski, A.~A.\ 1995, \mnras, 273, 837
\bibitem[Magorrian et al.(1998)]{Magorrian1998} Magorrian, J., et al.\ 1998, \aj, 115, 2285
\bibitem[Malizia et al.(2009)]{Malizia2009} Malizia, A., Stephen, J.~B., Bassani, L., Bird, A.~J., Panessa, F., \& Ubertini, P.\ 2009, \mnras, 399, 944
\bibitem[Marconi \& Hunt(2003)]{Marconi-Hunt2003} Marconi, A., \& Hunt, L.~K.\ 2003, \apjl, 589, L21
\bibitem[Matt et al.(1991)]{Matt1991} Matt, G., Perola, G.~C., \& Piro, L.\ 1991, \aap, 247, 25
\bibitem[Mitsuda et al.(2007)]{Mitsuda2007} Mitsuda, K., et al.\ 2007, \pasj, 59, 1
\bibitem[Mizuno et al.(2008)]{Mizuno2008} Mizuno, T., et al.\ 2008, JX-ISAS-SUZAKU-MEMO-2008-03
\bibitem[Moran et al.(2001)]{Moran2001} Moran, E.~C., Kay, L.~E., Davis, M., Filippenko, A.~V., \& Barth, A.~J.\ 2001, \apjl, 556, L75
\bibitem[Morganti et al.(1993)]{Morganti1993} Morganti, R., Killeen, N.~E.~B., \& Tadhunter, C.~N.\ 1993, \mnras, 263, 1023
\bibitem[Murphy \& Yaqoob(2009)]{Murphy2009} Murphy, K.~D., \& Yaqoob, T.\ 2009, \mnras, 397, 1549
\bibitem[Nakajima et al.(2008)]{Nakajima2008} Nakajima, H., et al.\ 2008, \pasj, 60, 1
\bibitem[Nandra \& Pounds(1994)]{Nandra1994} Nandra, K., \& Pounds, K.~A.\ 1994, \mnras, 268, 405
\bibitem[Noguchi et al.(2009)]{Noguchi2009} Noguchi, K., Terashima, Y., \& Awaki, H.\ 2009, \apj, 705, 454
\bibitem[Ozawa et al.(2009)]{Ozawa2009} Ozawa, M., et al.\ 2009, \pasj, 61, 1
\bibitem[Parisi et al.(2009)]{Parisi2009} Parisi, P., et al.\ 2009, \aap, 507, 1345
\bibitem[Risaliti et al.(2002)]{Risaliti2002} Risaliti, G., Elvis, M., \& Nicastro, F.\ 2002, \apj, 571, 234
\bibitem[Storchi-Bergmann et al.(1995)]{Storchi-Bergmann1995} Storchi-Bergmann, T., Kinney, A.~L., \& Challis, P.\ 1995, \apjs, 98, 103
\bibitem[Tueller et al.(2008)]{Tueller2008} Tueller, J., Mushotzky, R.~F., Barthelmy, S., Cannizzo, J.~K., Gehrels, N., Markwardt, C.~B., Skinner, G.~K., \& Winter, L.~M.\ 2008, \apj, 681, 113
\bibitem[Ueda et al.(2003)]{Ueda2003} Ueda, Y., Akiyama, M., Ohta, K., \& Miyaji, T.\ 2003, \apj, 598, 886
\bibitem[Ueda et al.(2007)]{Ueda2007} Ueda, Y., et al.\ 2007, \apjl, 664, L79
\bibitem[Verner et al.(1996)]{Verner1996} Verner, D.~A., Ferland, G.~J., Korista, K.~T., \& Yakovlev, D.~G.\ 1996, \apj, 465, 487
\bibitem[Winter et al.(2008)]{Winter2008} Winter, L.~M., Mushotzky, R.~F., Tueller, J., \& Markwardt, C.\ 2008, \apj, 674, 686
\bibitem[Winter et al.(2009)]{Winter2009} Winter, L.~M., Mushotzky, R.~F., Reynolds, C.~S., \& Tueller, J.\ 2009, \apj, 690, 1322
\end{thebibliography}
\end{document}